\documentclass[aps,nofootinbib,pra,amsmath,amssymb,twocolumn]{revtex4}

\usepackage{graphicx}
\usepackage{dcolumn}
\usepackage{bm}
\usepackage{etoolbox}

\usepackage[utf8]{inputenc}
\usepackage[T1]{fontenc}
\usepackage[english]{babel}

\usepackage{gensymb}
\usepackage{dsfont,amscd,mathrsfs,braket}
\usepackage[centercolon=true]{mathtools}
\usepackage{microtype,xspace,listings}
\usepackage[usenames,dvipsnames]{xcolor}

\usepackage{csquotes}
\usepackage{hyperref}
\usepackage[capitalise]{cleveref}
\usepackage{lipsum}
\usepackage{graphicx}
\usepackage{epsfig,amssymb,amsmath}

\newcommand*{\swap}[2]{%
  \let\temp#1 \let#1#2 \let#2\temp \let\temp\relax}
\swap{\epsilon}{\varepsilon}
\swap{\theta}{\vartheta}
\swap{\phi}{\varphi}

\hypersetup{%
  colorlinks=true,%
  pdfstartpage=1,%
  pdfstartview=FitV,%
  breaklinks=true,%
  pageanchor=true,%
  bookmarksnumbered,%
  bookmarksopen=true,%
  bookmarksopenlevel=1,%
  hypertexnames=true,%
  pdfhighlight=/O,%
  linkcolor=RoyalBlue,
  citecolor=red,%
}

\lstdefinelanguage{Julia}%
  {morekeywords={abstract,break,case,catch,const,continue,do,else,elseif,%
      end,export,false,for,function,immutable,import,importall,if,in,%
      macro,module,otherwise,quote,return,switch,true,try,type,typealias,%
      using,while},%
   sensitive=true,%
   morecomment=[l]\#,%
   morecomment=[n]{\#=}{=\#},%
   morestring=[s]{"}{"},%
}[keywords,comments,strings]

\newcommand*{\tp}[1]{_{\textup{#1}}}

\renewcommand*{\Re}{\mathrm{Re}}

\newcommand*{\ketbra}[2]{\ket{#1}\!\!\mkern2mu\bra{#2}}

\DeclarePairedDelimiter{\abs}{\lvert}{\rvert}

\DeclarePairedDelimiter{\pp}{\lparen}{\rparen}

\newcommand*{\deq}{\coloneqq}

\newcommand*{\dd}{\mathrm{d}\mkern0mu}
\newcommand*{\ct}[1]{#1^{\dagger}}

\newcommand*{\mmc}[1]{\mathcal{#1}}
\newcommand*{\cL}{\mmc L}
\newcommand*{\cG}{\mmc G}
\newcommand*{\msf}[1]{\mathsf{#1}}

\newcommand*{\numberset}{\mathbb}

\newcommand*{\R}{\numberset{R}}
\newcommand*{\C}{\numberset{C}}

\newcommand{\Tr}{\mathrm{Tr}}

\def\>{\rangle}
\def\<{\langle}

\def\commentg#1{\relax}

\def\togli#1{\relax}

\begin{document}

\preprint{APS/123-QED}

\title{High-Dimensional Methods for Quantum Homodyne Tomography}

\author{Nicola Mosco}
\email{nicola.mosco@unipv.it}
\author{Lorenzo Maccone}%
\email{maccone@unipv.it}
\affiliation{%
  {Dip.~Fisica and INFN Sez.\ Pavia, University of Pavia, via
    Bassi 6, I-27100 Pavia, Italy}
}

\date{\today}

\begin{abstract}
  We provide optimized recursion relations for homodyne
  tomography. We improve previous methods by mitigating the
  divergences intrinsic in the calculation of the pattern
  functions used previously, and detail how to implement the
  data analysis through Monte Carlo simulations. Our
  refinements are necessary for the reconstruction of
  excited quantum states which populate a high-dimensional
  subspace of the electromagnetic field Hilbert space. We
  also present a Julia package for the analysis and
  the reconstruction method.
\end{abstract}


\maketitle

Quantum tomography is the procedure that reconstructs the
quantum state of a system from repeated measurements of a
(complete) set of observables on a number of copies of
equally prepared quantum systems. This is the only way to
measure a quantum state: indeed, on one hand it is
impossible to recover the state from a single copy of the
system~\cite{mauro}; on the other, without the measurement
of a complete set of observables (a quorum), there is not
enough information for the reconstruction as different
states may give the exact same statistics on an incomplete
set of observables. The same quantum systems may have
different possible quorums~\cite{quorum}. For the state of a
single mode of the radiation field there are two quorums
that are typically used: either the field quadratures
$X_\phi = \frac{1}{2}(\ct a e^{i\phi} + a
e^{-i\phi})$~\cite{d2003quantum,DAriano2004,reviewtomo} or
the displaced parity operator $D^\dag(\alpha)(-1)^{a^\dag
a}D(\alpha)$~\cite{davidovich}, where $a$ is the
annihilation operator of the mode and $D(\alpha) = e^{\alpha
\ct a - \bar\alpha a}$ is the displacement operator. In this
paper we will focus on the former, which can be detected
straightforwardly with a homodyne
detector~\cite{reviewtomo}. The homodyne measurements are
the marginal probability distributions of each quadrature
$X_\phi$, collectively giving the Radon transform of the
Wigner function. As such, the seminal early homodyne
tomography method employed Radon transform inversion
techniques~\cite{vogel}. This method, thus, requires the
homodyne data to be binned in order to obtain a
distribution.  This introduces a bias in the reconstruction,
due to the width of the binning. The Homodyne Computed
Tomography (HCT) method, proposed in~\cite{chiara} and
refined in
\cite{paul1995measuring,PhysRevA.52.R1801,PhysRevA.52.4899},
avoids this problem by directly reconstructing the quantum
state---or expectation values of arbitrary
operators---without going through the Wigner function.
Hence, one could in principle use a single data point for
each homodyne measurement (no binning) to ensure complete
unbiasedness. Alternative procedures use maximum likelihood
methods~\cite{maxlik}, which produce very high-quality
reconstructions, but they are biased and, especially,
completely unsuited for the high-dimensional states that we
consider here: the minimization procedures entailed become
rapidly intractable for large dimensions.

In this paper we provide a (small) further refinement of the
HCT method to tame numerical instabilities, and present a
Julia package from which some extremely high-dimensional
Monte Carlo simulations of reconstructions are obtained to
illustrate the method. These show that quantum states with
density matrices of one order of magnitude larger than the
size of previous simulations and experiments can be
reconstructed.\togli{, attaining density matrices of
dimension $M\times M$ with up to $M \approx 2000$ using
single precision floats in the reconstruction algorithm and
$M \approx 4000$ using double precision numerics.}  This is
timely since quantum tomography is starting to be used in
microwave cavities \cite{science,wallr1,wallr,mallet}, where
even small signals entail huge number of photons, i.e.~a
high-dimensional subspace of the radiation Hilbert space
because of the low energy of each microwave photon.

The outline follows: in \cref{sec:homodyne-tomog} we provide
a bare-bones description of the method; in
\cref{sec:num-impl} we describe how one can produce a
numerical algorithm that implements it, and we illustrate it
with simulated experiments; in \cref{sec:software} we
describe the software packages used for this work; and,
finally, in \cref{sec:conclusions} we draw our conclusions
regarding the work presented in this paper. In
\cref{sec:review-tomo-formulae} we review the derivation of
the method and in \cref{sec:recur-wigner} we derive the
optimized recursion relations for the calculation of the
Wigner function.

\section{Homodyne Tomography} \label{sec:homodyne-tomog}

\begin{figure}[t]
  \includegraphics[width=0.5\textwidth]{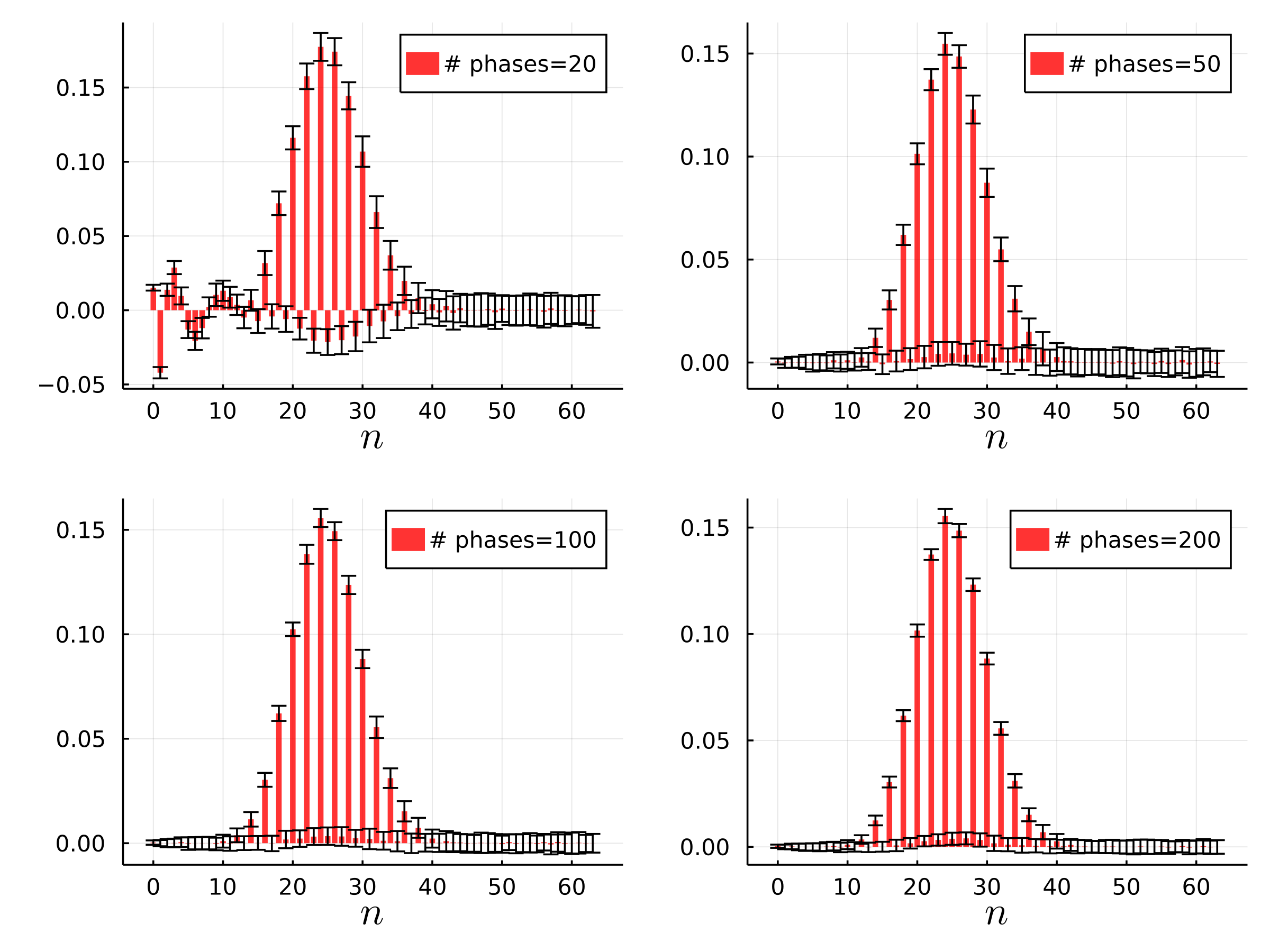}
  \caption{Effect of the number of phases: the pictured
    tomographic reconstructions (diagonal of the density
    matrix) use the same number of quadrature results
    $n\tp{bin} = 400$ for different values of the number of
    quadrature phases $n_\phi \in \{20,50,100,200\}$. The
    reconstructions rapidly converge to a good quality
    already for relatively small numbers of phases. The
    reconstructions refer to a Schrödinger cat state
    $\propto\ket{\alpha} + \ket{-\alpha}$ with $\alpha=5$
    and density matrix cut-off $M=64$. The data was
    generated with a Monte Carlo simulation with $1000$
    quadrature measurements and $100$ statistical blocks for
    a total of $1000 \times 100$ measurements for each
    phase.}
  \label{fig:nfasi}
\end{figure}

The homodyne detector is an apparatus that measures the
quadrature observable $X_\phi$ at optical frequencies.
Similar devices can be built for microwave cavities,
e.g.~\cite{wallr1,mallet}. This apparatus receives as input
a number $\phi \in [0,\pi]$, which typically refers to the
phase of a local oscillator in an intense coherent state. It
outputs a real number $x$, which (once calibrated) denotes
the measured value of the quadrature $X_\phi$ identified by
$\phi$. The data taking stage of the homodyne tomography
experiment then resorts in choosing $n_\phi$ values $\phi_i$
of the phase and measuring $n_q$ quadrature values $x_i$ for
each chosen phase. Ideally, one should choose $n_q=1$
(i.e.~use a different phase for each quadrature measurement)
to avoid a possible source of bias, but it is
known~\cite{phases}, and confirmed by our simulations
(e.g.~\cref{fig:nfasi}), that this is not a big concern if
the number of phases $n_\phi$ is sufficiently large (more
excited states requiring a larger number of phases). Using
fixed phases simplifies the experiment considerably. The
$n_\phi$ phases must be chosen uniformly in the interval
$[0,\pi]$ or, more conveniently for the numerical
implementation, in the interval $[0,2\pi]$: due to the phase
space symmetries, the two intervals are equivalent as
discussed below.

The result of the experiment is a set of data $\{(\phi_k,
x_k)\}_{k\in I_N}$, $I_N = \set{0,\dots,N-1}$, where $x_k$
is the result of the $k$-th measurement in which the
quadrature $X_{\phi_k}$ was measured and $N=n_\phi n_q$ is
the total number of measurements. \togli{$k=0,\cdots,N-1$,
where $x_k$ is the result of the $k$-th measurement in which
the quadrature $X_{\phi_k}$ was measured and $N = n_\phi
n_q$ is the total number of measurements.}

Now we detail how this data can be used for the tomographic
reconstruction. Importantly, the HCT method allows for the
direct reconstruction of the expectation value $\braket{O}$
of \emph{any} operator, not necessarily an observable. As an
illustrative example, we now show how one can reconstruct
the density matrix $\rho_{nm}=\braket{n|\rho|m}$ written on
the Fock basis. This is achieved by using the non-Hermitian
$O=\ketbra{m}{n}$, since $\braket{O} =
\Tr[\rho\ketbra{m}{n}] = \braket{n|\rho|m}$. It can be shown
(see Appendix, \cref{eq:rho-tomo}) that the matrix element
can be written as
\begin{align}
  \rho_{n,m} = \frac{1}{\pi} \int_0^{\pi} \!\!\!
    \dd\phi \, e^{-i(m-n)\phi} \int_R \dd x \,
    p_\phi(x) f_{n,m}(x), \label{eq:rho-tomo1}
\end{align}
where $f_{n,m}$ are pattern functions defined below and
$p_\phi(x) = {}_\phi \! \braket{x|\rho|x}_\phi$ is the
conditional probability of obtaining outcome $x$ given that
the $\phi$-quadrature was measured, namely the outcome
relative to the eigenstate $\ket{x}_\phi$ of $X_\phi$.
Since, by construction, the phases $\phi_i$ are uniform in
the interval $[0,\pi]$, we can interpret $1/\pi$ as the
uniform probability of choosing the phase $\phi$; then, from
Bayes rule, the joint probability of choosing a phase $\phi$
\emph{and} obtaining a $\phi$-quadrature outcome $x$ is
$p_{\phi}(x)/\pi$.  Thus we can use the collected data
$\{(\phi_k,x_k)\}_{k\in I_N}$ to calculate the double
integral of \cref{eq:rho-tomo1} as a Monte Carlo integral:
\begin{align}
  \rho_{n,m} = \frac 1N \sum_{k=0}^{N-1}
    e^{-i(m-n)\phi_k} f_{n,m}(x_k)
    \pm \Delta_N, \label{mc}
\end{align}
where $\Delta_N$ is the error due to the finite $N$. Indeed,
the equality is strict only in the limit $N\to\infty$ (when
$\Delta_N\to 0$), which of course cannot be reached
experimentally. However, since we are using Monte Carlo
integration, the error $\Delta_N$ for finite $N$ is
\emph{strictly a statistical error only}, which can be
easily evaluated from the variance $\sigma^2$ of the data:
\begin{multline}
  \Delta_N =
  \sqrt{\frac{\sigma^2}{N}} = \\
  \sqrt{\frac{1}{N(N-1)}
  \left[\textstyle\sum_k F^2(\phi_k,x_k) -
  \big(\textstyle\sum_k F(\phi_k,x_k)\big)^2\right]},
\end{multline}
with $F(\phi_k,x_k)=e^{-i(m-n)\phi_k}f_{n,m}(x_k)$ (more
rigorously, if $m\neq n$, when $F$ is complex, one has to
separately evaluate the variance of the real and imaginary
parts).

\section{Numerical implementation} \label{sec:num-impl}

The calculation of the pattern functions
\cite{leorichter,PhysRevA.52.4899} is quite involved and is
reviewed in \cref{sec:review-tomo-formulae}. Starting from
the recursion relations derived in the same section, we
provide here a simplified form, suitable for numerical
evaluation also with non-wide floating point representation,
such as the standard 32 bits single precision data type.

\begin{figure}[t]
  \centering
  \includegraphics[width=0.5\textwidth]{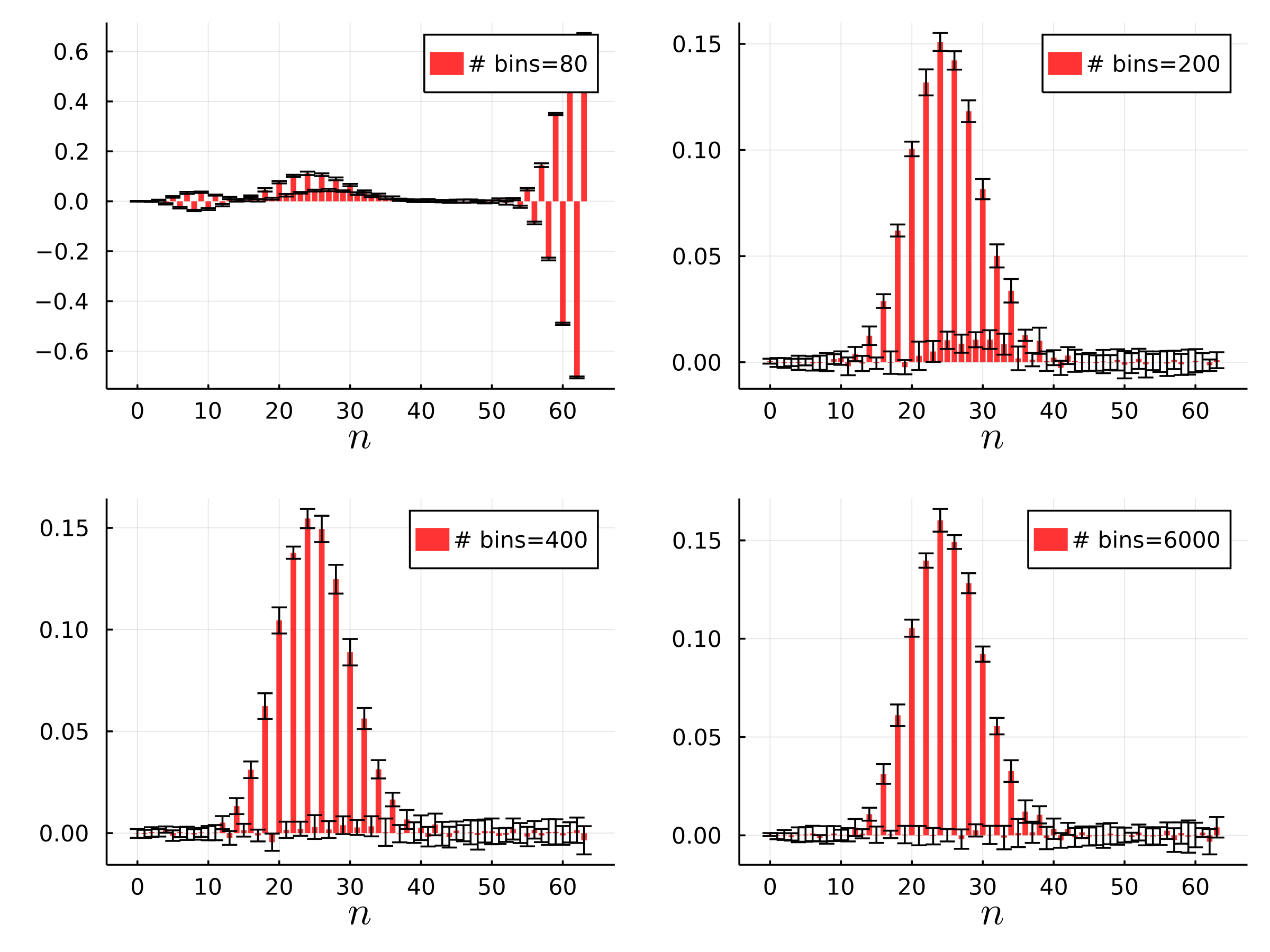}
  \caption{Effect of the data binning. While HCT does not
    require binning of the homodyne data, it can speed up
    the data analysis. We show the diagonal of the density
    matrix with error bars on each element, showing the
    effect of binning for various bin size: the
    reconstruction converges quickly when increasing the
    number of bins $n\tp{bin}$ and gives good results also
    in the regime $n\tp{bin} \gg 1$, where the inverse Radon
    transform method fails. The measurements were obtained
    with a Monte Carlo simulation of $100$ measurements for
    $10$ statistical blocks with $n_\phi=800$ phases. The
    simulated state is a Schrödinger cat state $\ket{\psi}
    \propto \ket{\alpha} + \ket{-\alpha}$, $\alpha=5$ and
    density matrix dimension $M=64$.}
  \label{fig:bins}
\end{figure}

To speed up the data analysis, we create a probability
distribution by discretizing the possible values of the
quadrature $x$ and binning the $n_q$ values of the
quadrature data $x_k$ into $n\tp{bin}$ bins. As discussed
above, in contrast to methods based on the inversion of the
Radon transform, this is not a fundamental limitation.
Indeed, we show in \cref{fig:bins} that our method still
works in the regime where $n\tp{bin}\gg n_q$, where most
bins are unpopulated and at most a handful of data points
end up in the same bin. In this regime, the Radon transform
inversion would fail, in contrast to HCT. Nonetheless,
binning may be useful, since it allows for a faster and more
effective reconstruction because one has to calculate the
recurrence relation below only once per bin (this is of
interest for the reconstruction of huge data
sets).\togli{The input data are assumed to be a discrete
collection of homodyne measurements $\msf S_N =
\set{(\phi_k,x_k) | k\in I_{N}}$, with $I_N =
\set{0,\dots,N-1}$. This collection represents a discrete
sampling of the probability distribution $p_\phi(x) =
{}_\phi\!\braket{x|\rho|x}_\phi$.}

The binning is done by collecting the data into a $n_\phi
\times n\tp{bin}$ matrix $S$, called \emph{sinogram}. Each
matrix row $j\in I_{n_\phi}$ represents the quadrature
probability at phase $\phi_j$, namely $S_{j,i}$ is the
fraction of the $n_q$ values of $x_k$ that fall into the
$i$-th bin, $i\in I_{n\tp{bin}}$.\togli{, after having
introduced $n\tp{bin}$ bins for the homodyne data $x_k$.}
Asymptotically we have that $S_{j,i} \sim
p_{\phi_j}(x_i)$, with $x_i$ a representative point in the
$i$-th bin. We choose equispaced phases $\phi_j =
\frac{2\pi}{n_\phi} j$. Again, this is a matter of numerical
convenience: in principle, one should choose the phases with
uniform random probability in the interval $[0,2\pi]$. We
assume the measurements are taken on the whole interval
$[0,2\pi]$ so that we can use the FFT algorithm; if this is
not the case---as it is when the samples are taken only in
$[0,\pi]$---one can always double the data exploiting the
phase space symmetry $X_{\phi+\pi} = -X_{\phi}$.
\togli{Hence, the input data $\msf S_N$ can be collected
into a real matrix $S_{j,i}$ with dimensions $n_\phi \times
n\tp{bin}$.}

In order to implement the tomographic formula on a computer
we have to choose a cutoff for the density matrix. As will
be clear in the following, this cutoff does not introduce
any bias in the reconstruction, although a sensible
(i.e.~normalized) reconstruction will be obtained only if
such cutoff is chosen sufficiently large. Indeed, one can
check a posteriori whether the chosen $M$ is sufficient by
looking at the normalization of the reconstructed state,
namely by checking if it is compatible with one within the
statistical error bars.  In the following we will assume the
density matrix has dimensions $M\times M$.

By dividing the sum into two parts (along the bins and along
the phases), the Monte Carlo integral of \cref{mc} becomes
\begin{align}
  & \rho_{n,m} = \sum_{i=0}^{n\tp{bin}-1}
    \hat S_{m-n,i} f_{n,m}(x_i),
\intertext{where the factor $1/n\tp{bin}$ is included into the
  definition of $\hat S$ and}
  & \hat S_{d,i} = \frac{1}{n_\phi} \sum_{j=0}^{n_\phi-1}
    S_{j,i} e^{-2\pi i\frac{jd}{n_\phi}}
\end{align}
is the one-dimensional discrete Fourier transform (DFT)
along the first dimension of $S_{j,i}$.

\begin{figure}[t]
  \centering
  \includegraphics[width=0.4\textwidth]{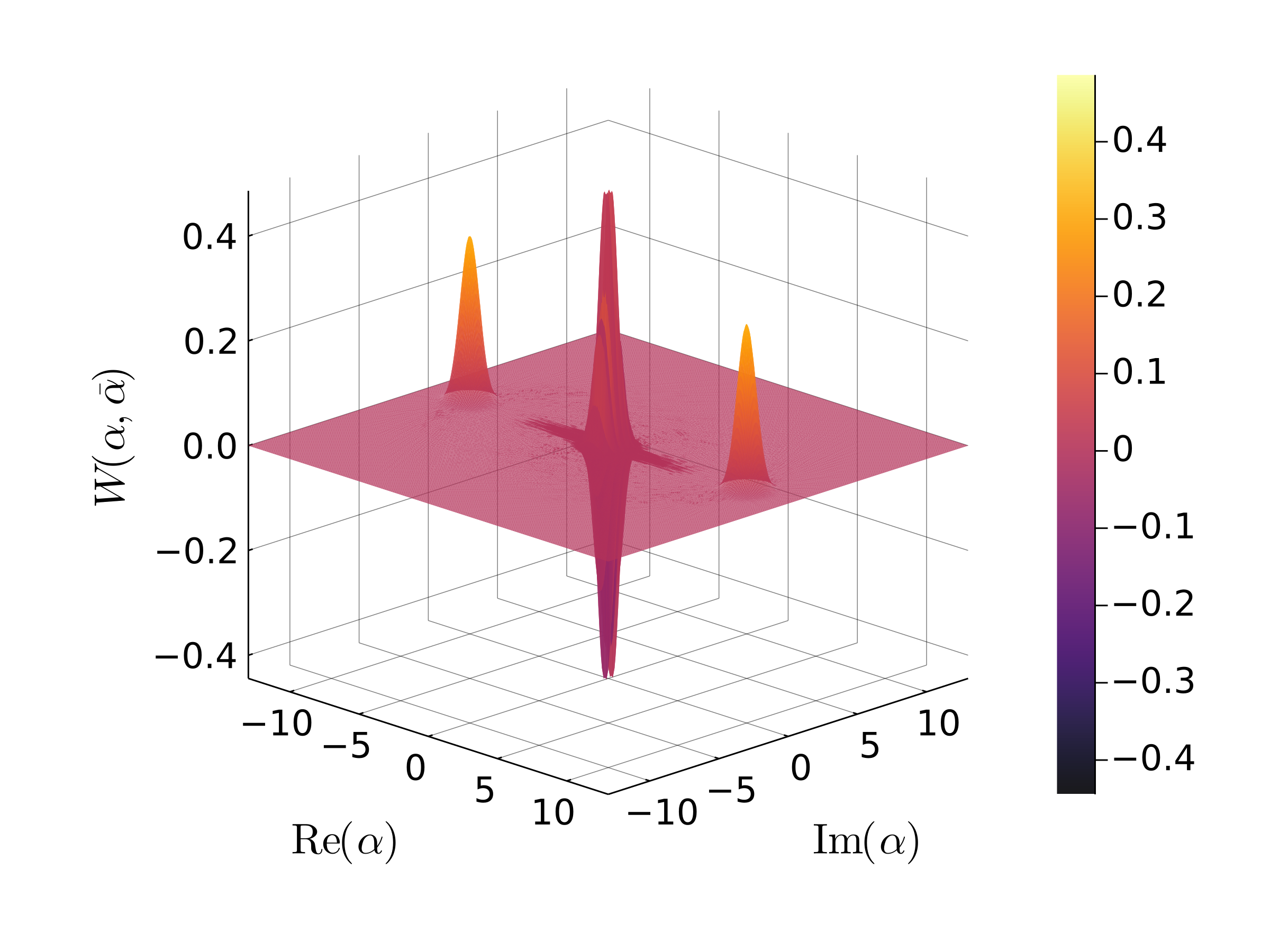}
  \caption{Wigner function $W(\alpha,\bar\alpha)$ for a
  Schrödinger cat state $\ket{\psi} \propto \ket{\alpha} +
  \ket{-\alpha}$, $\alpha=10$, $n\tp{bin} = 1000$ and
  $M=256$; number of samples: $10000$ for $100$ statistical
  blocks and $n_\phi=800$ phases.}
  \label{fig:wig-cat}
\end{figure}

The pattern functions $f_{n,m}$ can be written in the
following factorized
form~\cite{PhysRevA.52.R1801,PhysRevA.52.4899,leorichter}
(see \cref{sec:review-tomo-formulae}):
\begin{multline} \label{eq:pf-recursion1}
  f_{n,m}(x) = 2x u_n(x) v_m(x) - \phantom{x} \\
    \sqrt{n+1} u_{n+1}(x) v_m(x) - \phantom{x} \\
    \sqrt{m+1} u_{n}(x) v_{m+1}(x),
\end{multline}
The vectors $u_n$ and $v_m$ appear always multiplied
together, which means that we can change their definition
with respect to their previously published values
\cite{leorichter,PhysRevA.52.4899} (see
\cref{eq:u-rec,eq:v-rec-b,eq:v-rec-f}) for the purposes of
the numerical implementation. Furthermore, the square roots
$\sqrt{n}$ appear always with the same index of either $u_n$
or $v_m$, so we introduce additional vectors $\tilde u_n =
\sqrt{n} u_n$, $\tilde v_m = \sqrt{m} v_m$ that can be
precomputed along with $u_n$ and $v_m$. Thus, a recursive
definition of $u_n$ that is better suited for numerics is
\begin{equation}
\begin{cases}
  & u_0 = \beta, \quad \tilde u_0 = 0, \\
  & u_1 = 2x u_0, \quad \tilde u_1 = u_1, \\
  & \tilde u_n = 2x u_{n-1} - \sqrt{n-2} u_{n-2}, \\
  & u_n = \tilde u_n / \sqrt{n}.
\end{cases}
\end{equation}
Here, we introduced a constant $\beta$ to compensate for growth of the
sequence $(u_n)$ and can be arbitrarily chosen to control its range of
values. The simulations and reconstruction presented here are obtained
with the heuristic choice $\beta \approx \exp(-3\max x)$, where $\max
x$ is the maximum obtained homodyne value.

Likewise, the corresponding definition of $v_m$ best suited
for numerics follows. It is defined through a forward or
backward recursion \cite{leorichter,PhysRevA.52.4899}
depending on the value of $x$. The backward recursion can be
used if
\begin{align} \label{eq:safe-region1}
  \abs{x} < \alpha_{4k} -
    \frac{1}{2}\frac{1}{\sqrt[3]{\alpha_{4k}}},
\end{align}
with $\alpha_n = \sqrt{n+1/2}$. Then the recursion can be
safely started from $m=4M$:
\begin{equation}
\begin{cases}
  & v_{4M}(x) = \beta^{-1} e^{-x^2} \kappa_{4M}, \\
  & \tilde v_{4M}(x) = \sqrt{4M} v_{4M}(x), \\
  & v_{4M-1}(x) = \beta^{-1} e^{-x^2} \kappa_{4M-1}, \\
  & \tilde v_{4M-1}(x) = \sqrt{4M-1} v_{4M-1}(x), \\
  & v_m(x) = \frac{1}{\sqrt{m+1}}
    (2x v_{m+1} - \sqrt{m+2} v_{m+2}), \\
  & \tilde v_m(x) = \sqrt{m} v_m(x).
\end{cases}
\end{equation}
For values of $x$ outside the region of
\cref{eq:safe-region1}, one should use the forward recursion
\begin{equation}
\begin{cases}
  & v_0(x) = \frac{\beta^{-1}}{x}, \\
  & \tilde v_0(x) = 0, \\
  & v_m(x) = \frac{\sqrt{m}}{2x} v_{m-1}(x), \\
  & \tilde v_m(x) = \sqrt{m} v_m(x).
\end{cases}
\end{equation}
We scaled the starting value by $\beta^{-1} e^{-x^2}$ to
compensate the divergence of the vectors $v_m$ and also
canceling the effect of the constant $\beta$, which is
needed only to change the range of values assumed by the
vectors $u_n$ and $v_m$. Now the pattern functions $f_{n,m}$
can be rewritten as
\begin{multline}
  f_{n,m}(x) =
    [2x u_n(x) - \tilde u_{n+1}(x)] v_m(x) - \phantom{x} \\
      u_n(x) \tilde v_{m+1}(x).
\end{multline}

\begin{figure}[t]
  \centering
  \includegraphics[width=0.5\textwidth]{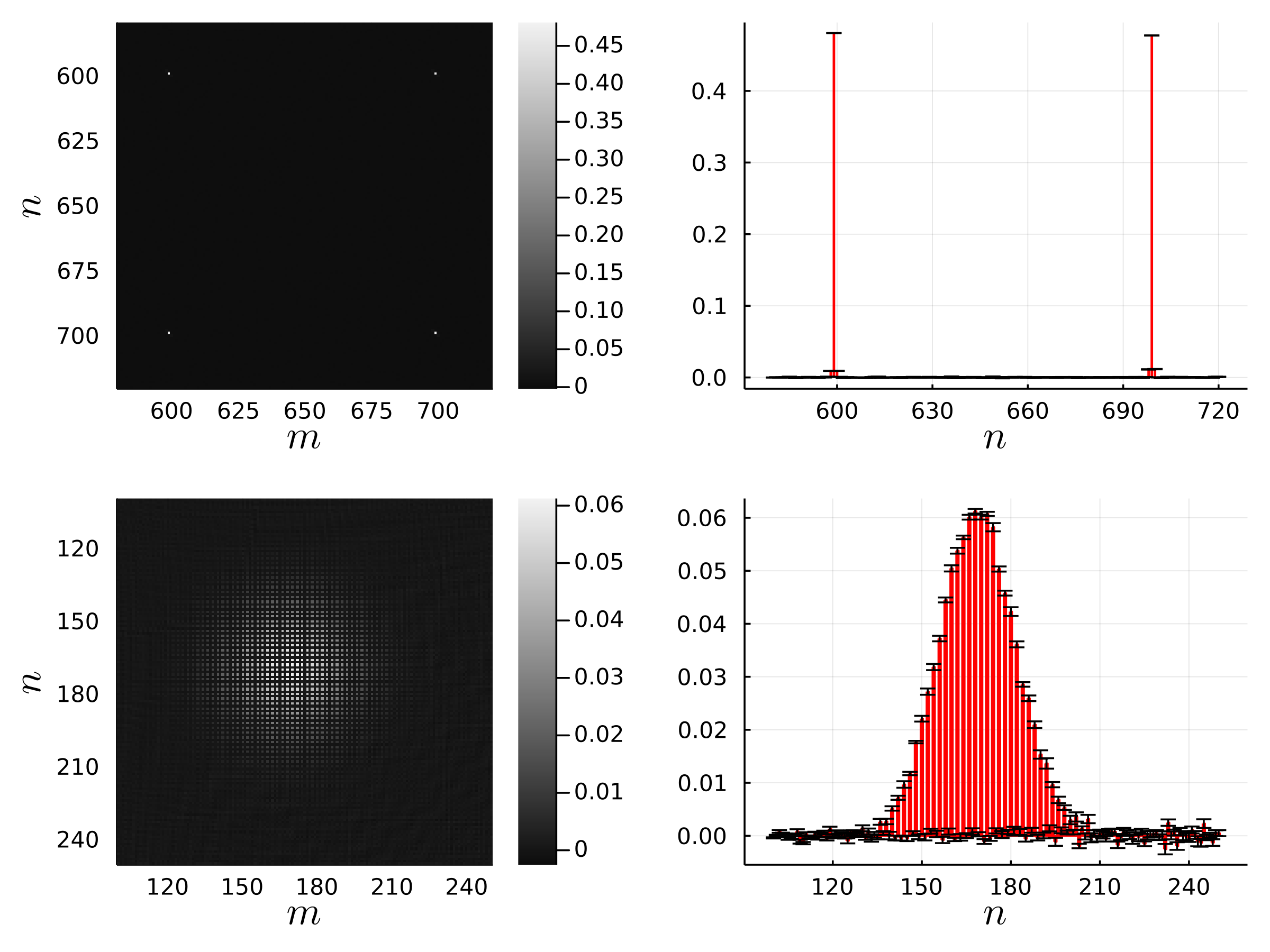}
  \caption{Simulated homodyne reconstructions of large density
    matrices. Left: the real part of the density matrix is plotted.
    Right: the diagonal is shown with the error bars on each matrix
    element. The plots above refer to the state
    $\ket{\psi} \propto \ket{600} + \ket{700}$, with $1000$ samples
    for $10$ blocks, $8000$ bins and $M=800$; the plots below to the
    state $\ket{\psi'} \propto \ket{\alpha} + \ket{-\alpha}$, with
    $\alpha=13$, $M=300$, $6000$ bins, and same statistics. A very small effect of
    the binning is evident as a slight decrease of the peaks in the
    upper-right figure.}
  \label{fig:result}
\end{figure}

In Fig.~\ref{fig:result} some examples of very large
reconstructions of density matrices are presented, which use
the above recursions.

\subsection{Calculating the Wigner function}

As discussed above, the HCT method does not require the
reconstruction of the Wigner function, as it directly
reconstructs the density matrix (and the expectation value
of arbitrary operators). Still, it may be useful to
calculate also the Wigner
function~\cite{Dariano1997Bilkent}, defined as
\begin{align}
  W(\alpha,\bar\alpha) = \frac{2}{\pi}
    \Tr\left[\rho e^{2\pp{\alpha \ct a - \bar\alpha a}}
      e^{i\pi \ct a a}\right],
\end{align}
which can be explicitly written as
\begin{align}
  & W(\alpha,\bar\alpha) = \frac{2}{\pi}
    \Re\sum_{d=0}^{+\infty} e^{id\arg{\alpha}}
      \sum_{n=0}^{+\infty}
      \Lambda_{n,d}(\abs{2\alpha}^2) \rho_{n,n+d},
      \label{eq:wigner-lambda} \\
\intertext{where $\rho_{n,m} = \braket{n|\rho|m}$, and}
  & \Lambda_{n,d}(x) = (-1)^n (2-\delta_{d0})
    x^{d/2} \sqrt{\frac{n!}{(n+d)!}} \cL_n^d(x).
      \label{eq:lambda-def}
\end{align}
Here, $\cL_n^d$ is defined as $\cL_n^d(x) = e^{-x/2}
L_n^d(x)$, where $L_n^d(x)$ are the (generalized) Laguerre
polynomials.

The formula in \cref{eq:lambda-def} does not perform well
for lower precision data types as there are several
quantities that become very large quite fast. Although one
can mitigate this by, for instance, computing first the
logarithm of the factorials, this may still be not enough
for highly excited states that occupy a large part of the
phase space. This problem can be solved with appropriately
designed recursion relations that we present in
\cref{sec:recur-wigner}. The two methods presented there are
inequivalent, but have roughly the same efficiency.

\begin{figure}[t]
  \centering
  \includegraphics[width=0.5\textwidth]{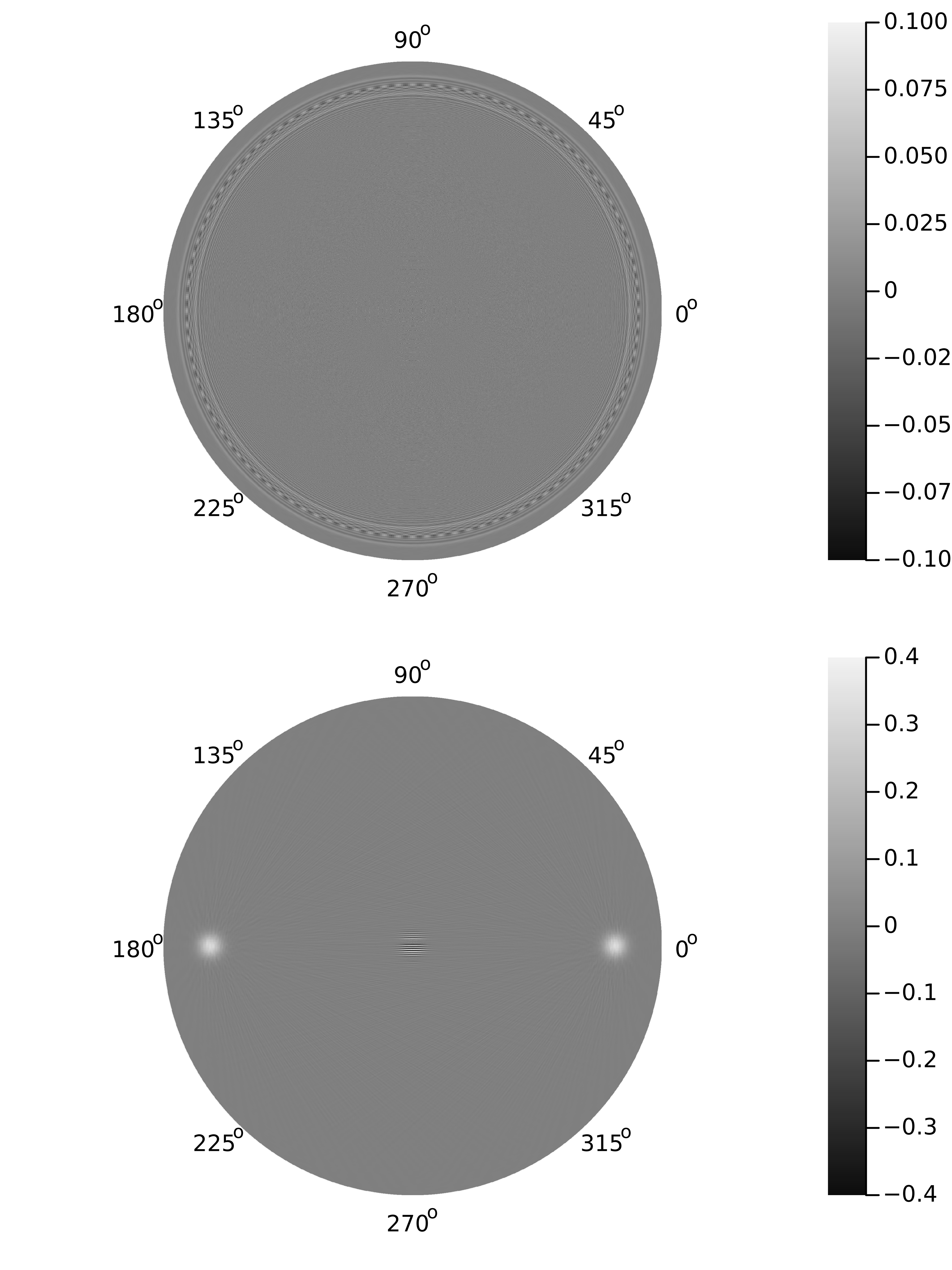}
  \caption{Wigner functions of the simulations presented in
    \cref{fig:result}. Top: $\ket{\psi} \propto \ket{600} +
    \ket{700}$; bottom: $\ket{\psi'} \propto \ket{\alpha} +
    \ket{-\alpha}$, with $\alpha=13$. To calculate these,
    the optimized formula in \cref{eq:wigner-numerics} was
    used. A window has been applied to the images for a
    better contrast.}
  \label{fig:wigner}
\end{figure}

For the numerical implementation we write the Wigner
function in polar coordinates $W(r,\theta)$, with $\alpha =
re^{i\theta}$:
\begin{equation} \label{eq:wigner-numerics}
  W(r,\theta) = \Re \sum_{d=0}^{M-1}
    \frac{e^{i\theta d}}{1+\delta_{d,0}}
    \sum_{n=0}^{M-d-1} \lambda_{n,d}(4r^2) \tilde\rho_{n,d},
\end{equation}
where $\tilde\rho_{n,d} = (-1)^n \rho_{n,n+d}$ and
$\lambda_{n,d}$ can be defined recursively leveraging the
definition of the Laguerre polynomials, as shown in
\cref{sec:recur-wigner}. It is convenient to rewrite the
density matrix by diagonals, reorganizing its elements for
efficient memory access.

\section{Software package} \label{sec:software}

The algorithms presented in this paper have been implemented
in a Julia~\cite{bezanson2012julia,bezanson2017julia,julia}
package and a C++ package. The repositories are hosted on
GitLab under the HomodyneCT group~\cite{homodyne-ct}. The
C++ package documentation can be found
at~\cite{hct-tools-wiki}.

The Julia package ecosystem for imaging and quantum
tomography under the GitLab group HomodyneCT consists of two
packages: MartaCT~\cite{martact} and
HomodyneImaging~\cite{homodyne-imaging-wiki}. The first one
implements the tools needed for image analysis and the
traditional reconstruction algorithms with FBP methods (this
package is open-source and distributed under the MIT
license); the second one implements all the quantum
algorithms described in this paper (this package is
currently closed source, so if you need access to it, please
contact the authors).

The main implementation resides in the Julia package
HomodyneImaging, along with the companion package MartaCT.
The basic interface is defined in MartaCT, while
HomodyneImaging extends it, leveraging the \emph{multiple
dispatch} system of the language.




\begin{lstlisting}[float,caption={Code example for \cref{fig:example}.},label=lst:example]
  using HomodyneImaging, QuantumOptics, IntervalSets, Plots
  using MartaCT.Simulations
  M = 64 # density matrix dimension
  bs = FockBasis(M-1)
  a = 3
  norm = sqrt(2*(1+exp(-2*abs2(a))))
  psi = (coherentstate(bs, a) + coherentstate(bs, -a)) / norm
  rho = tensor(psi, psi')
  xs = linspace(-5..5, 600)
  W = wigner(rho, xs, xs, FFTWigner())
  fs = linspace(ORI(0..2pi), 300)
  marg = radon(W, xs, fs, RadonSquare())
  sim = SinogramHomodyneSimulation(
    nsamples=1000, nblks=10, ndim=M)
  rhosim, rhodiag, yerrors = simulate(sim, marg, xs, fs)
  Wsim = wigner(rhosim, xs, xs, FFTWigner())
  dmp = bar(0:M-1, rhodiag;
      c=:red,
      yerrors,
      leg=:none,
      lw=0,
      lc=:red,
      xguide="\$n\$",
      guidefontsize=14,
  )
  wigp = heatmap(xs, xs, Wsim;
      xguide="\$x\$",
      yguide="\$y\$",
      guidefontsize=14,
  )
  plot(
      dmp,
      wigp;
      dpi=300
  )
\end{lstlisting}

\begin{figure}[b]
  \centering
  \includegraphics[width=0.5\textwidth]{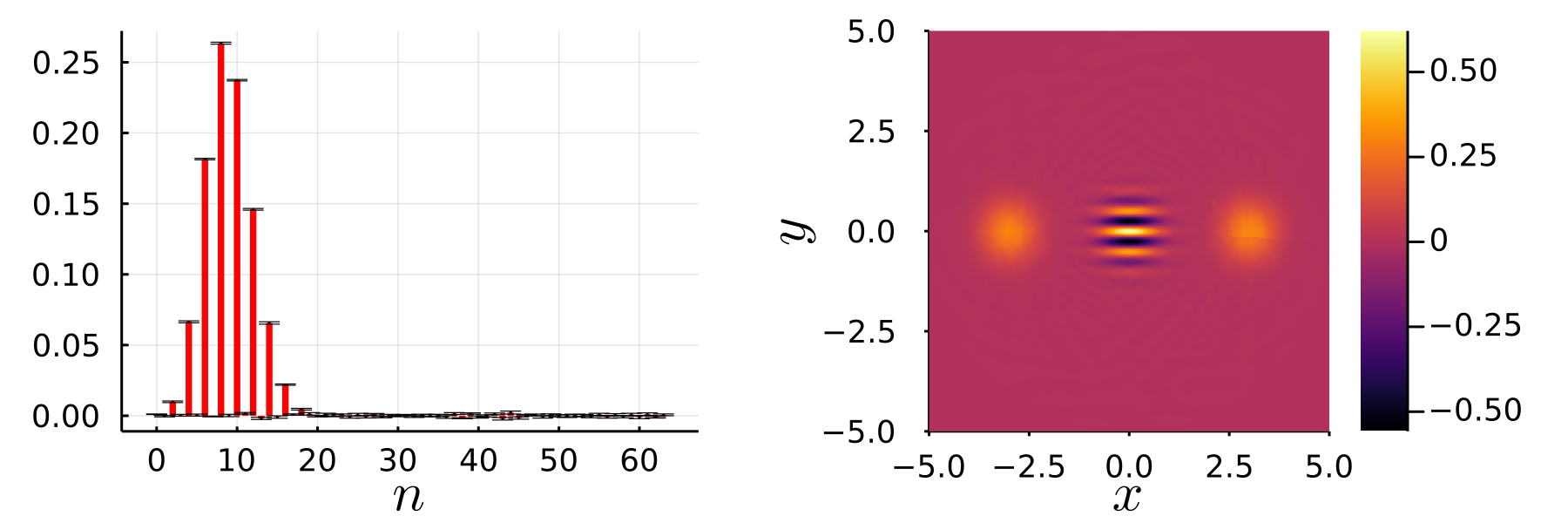}
  \caption{Example image obtained with the HomodyneImaging
  package representing a cat state $\ket{\psi} \propto
  \ket{\alpha} + \ket{-\alpha}$, with $\alpha=3$.}
  \label{fig:example}
\end{figure}

In order to give a sense to the reader of how MartaCT can be
used, we provide an example to reconstruct a symmetric cat
state $\ket{\psi} \propto \ket{\alpha} + \ket{-\alpha}$,
with $\alpha = 3$. The example \cref{lst:example} uses the
Julia package QuantumOptics~\cite{kramer2018quantumoptics}
to construct the quantum state. The resulting reconstruction
is depicted in \cref{fig:example}.

Let us dive in a bit in the source code of the example.
First of all, in Julia one needs to import the necessary
modules providing the funcionality we are going to use: this
is done with the \texttt{using} keyword, followed by a
comma-separated list of module names. We import several
modules: \texttt{HomodyneImaging} for the HCT algorithm;
\texttt{QuantumOptics} for the basic definitions regarding
quantum states; \texttt{IntervalSets} which provides a nice
syntax for intervals; \texttt{Plots} which is one of the
most used plotting packages available in Julia; and finally
we explicitly import \texttt{MartaCT.Simulations} which is a
sub-module providing the definition of the \texttt{simulate}
function which is used to perform a simulated experiment.

The parameters for this example are provided right after the
imported modules: we define a variable \texttt{M} which is
the dimension of the density matrix; the \texttt{bs}
variable that holds the basis data for the given dimension;
\texttt{a} the parameter of the coherent states. Then we
define the state \texttt{psi} as a symmetric superposition
of coherent states and the density matrix \texttt{rho}. The
Wigner function \texttt{W} is computed with the function
\texttt{wigner} using the algorithm \texttt{FFTWigner}
described in this paper.

In order to perform the simulation we need to compute the
marginals of the Wigner function first: this is achieved
computing the Radon transform of \texttt{W} with
\texttt{radon}, computing it on a square.

The simulation parameters are specified creating an object
of type \texttt{SinogramHomodyneSimulation}, whose name
recall the fact that we are providing the
marginals---namely, the sinogram. The result of the
simulation is stored in the variable \texttt{rhosim},
providing the simulated density matrix, along with the data
of the diagonal and its statistical errors. With the
simulated data we can compute again the Wigner function. The
example is concluded by the generation of the plots shown in
\cref{fig:example}.

\section{Conclusions} \label{sec:conclusions}

In conclusion, we have reviewed the homodyne tomography
technique, showing how to adapt it to high-dimensional state
reconstructions. We have described how this method can be
implemented in practice. We gave some illustrative examples
in the form of Monte Carlo simulations of the reconstruction
experiments. They demonstrate the robustness of the method.

Furthermore, we present the software packages implementing
the reconstruction algorithms. The main development has been
focused on the Julia packages providing a new promising
ecosystem for both medical imaging and quantum tomography
applications.

\section{Acknowledgments}

This work has been possible thanks to the support of the
\emph{ATTRACT} project ``Quantum Imaging for Tomography''
(QuIT)
\url{https://attract-eu.com/showroom/project/quantum-imaging-for-tomography-quit}
and also the \emph{Unitary Fund} project
\url{https://unitary.fund/grants.html}. LM acknowledges
support from the U.S. Department of Energy, Office of
Science, National Quantum Information Science Research
Centers, Superconducting Quantum Materials and Systems
Center (SQMS) under contract number DE-AC02-07CH11359.

\appendix

\section{Review of the derivation of the tomographic formulae}
\label{sec:review-tomo-formulae}

The tomographic reconstruction for the homodyne detector
relies on the fact that the displacement operators
$D(\alpha) = e^{\alpha \ct a - \bar\alpha a}$ are a complete
orthonormal basis for the space of operators with respect to
the Hilbert--Schmidt scalar product: $\braket{A|B} = \Tr[\ct
A B]$. More precisely, for any linear operator $A$ one can
write
\begin{equation} \label{eq:hs-disp-rep}
  \begin{split}
    A & = \frac{1}{\pi} \int_\C
      \Tr[A \ct D(\alpha)] D(\alpha) \, \dd^2 \alpha \\
      & = \frac{1}{4\pi} \int_0^\pi \int_\R
        \Tr[A e^{ikX_\phi}] e^{-ikX_\phi} \abs{k} \,
        \dd k \dd \phi,
    \end{split}
  \end{equation}
where $\alpha$ is written in polar coordinates in the last
equation: $\alpha = ike^{i\phi}/2$, and the first integral
can be taken only on the interval $[0,\pi]$ because of the
symmetry $X_{\phi+\pi} = -X_\phi$. The tomographic formula
is then obtained by introducing the probability $p_\phi(x) =
{}_\phi\!\braket{x|\rho|x}_\phi$ of getting $x$ when
measuring $X_\phi$:
\begin{align} \label{eq:tomo-formula}
  \braket{A} = \Tr[A\rho] =
    \frac{1}{\pi} \int_0^\pi \int_\R
    p_\phi(x) K_\phi[A](x) \, \dd x \dd\phi.
\end{align}
The operator $K_\phi[A]$ is the so-called \emph{kernel} of
homodyne tomography, defined by
\begin{align}
  K_\phi[A](x) = \frac{1}{4} \int_\R
    \Tr\left[Ae^{ik(X_\phi-x)}\right]
    \abs{k} \, \dd k.
\end{align}
Now, thanks to the trace in the integral, the kernel does
not necessarily diverge. It is possible to classify the set
operators $A$ which produce a bounded
kernel~\cite{Dariano02tomographicmethods} in
\cref{eq:tomo-formula}. In the simplest case considered
here, the operator $A$ should be at least Hilbert--Schmidt.
In particular this is the case of $\ketbra{n}{m}$ which
provides the matrix elements $\rho_{nm}$.

Since the first appearance~\cite{chiara} of this
method there have been several attempts to simplify the
expression of the tomographic formula for $\rho_{n,m}$. Let
us rewrite \cref{eq:tomo-formula} for $\rho_{n,m}$ with
explicit phase dependence:
\begin{align} \label{eq:rho-tomo}
  \rho_{n,n+d} = \frac{1}{\pi} \int_0^{\pi} \!\!\!
    \dd\phi \, e^{-id\phi} \int_R \dd x \,
    p_\phi(x) f_{n,n+d}(x),
\end{align}
where $f_{n,m}(x)$ are the so-called \emph{pattern
functions}, which are real and satisfy the symmetries:
\begin{align}
  & f_{n,m}(x) = f_{m,n}(x), \\
  & f_{n,m}(-x) = (-1)^{n-m} f_{n,m}(x).
\end{align}
The analytical expression for the pattern
functions$f_{n,m}(x)$ has been obtained by D'Ariano,
Leonhardt and
Paul~\cite{paul1995measuring,PhysRevA.52.R1801,PhysRevA.52.4899}
symplifying previous derivations as follows (notation and
conventions from~\cite{Dariano1997Bilkent}):
\begin{multline} \label{eq:kernel-rho}
  f_{n,m}(x) = 2 \sqrt{\frac{n!}{(n+d)!}}
    e^{-x^2} \cdot \phantom{x} \\
    \sum_{\nu=0}^n \frac{(-1)^\nu}{\nu!}
    \binom{n+d}{n-\nu} (2\nu+d+1)! \cdot \phantom{x} \\
    \Re\left[(-i)^d D_{-(2\nu+d+2)}(-2ix)\right],
\end{multline}
where $D_\lambda(z)$ denotes the parabolic cylinder
function. Such expression is very delicate to be used in
numerical implementations for large quantum numbers.

Further simplifications were possible thanks to an insight
of Richter~\cite{RICHTER1996327} who has been able to link
the tomographic formula to the regular and irregular
solutions of the Schrödinger equation of the harmonic
oscillator. The pattern functions are related to the
solutions of the Schrödinger equations employing an Hilbert
transformation. The pattern functions $f_{n,m}$ can be
written as a derivative of some functions
$g_{n,m}$~\cite{PhysRevA.52.4899,leorichter}:
\begin{align}
  f_{n,m}(x) = \frac{\partial g_{n,m}}{\partial x}(x),
\end{align}
where the functions $g_{n,m}$ are obtained through a Hilbert
transformation:
\begin{align}
  g_{n,m}(x) = \frac{\mathcal{P}}{\pi}
    \int_\R \frac{u_n(\xi)u_m(\xi)}{x-\xi} \, \dd \xi,
\end{align}
where $\mathcal{P}$ denotes the Cauchy principal value and
$u_n$ are the normalizable solutions of the Schrödinger
equation
\begin{align}
  \pp*{-\frac{1}{2}\frac{\partial^2}{\partial x^2}
    + \frac{x^2}{2}} u_n(x) =
      \pp*{n+\frac{1}{2}} u_n(x).
\end{align}
In the same ref.~\cite{leorichter}, the authors were
able to obtain an explicit expression for the functions
$g_{n,m}$:
\begin{align}
  g_{n,m}(x) = u_n(x) v_m(x), \quad
    \text{for $m \geq  n$,}
\end{align}
while for $m < n$ one can use $g_{n,m} = g_{m,n}$. Here, the
functions $v_m$ are the irregular solutions of the
Schrödinger equation, namely unnormalizable solutions
continued to the complex plane:
\begin{align}
  \pp*{-\frac{1}{2}\frac{\partial^2}{\partial z^2}
    + \frac{z^2}{2}} v_m(z) =
      \pp*{m+\frac{1}{2}} v_m(z),
\end{align}
with $v_m(z)$ real on the real axis $z = x$.

The evaluation of the pattern functions $f_{nm}$ can be
implemented efficiently on a computer by employing the
factorized form
\begin{multline} \label{eq:pf-recursion}
  f_{n,m}(x) = 2x u_n(x) v_m(x) - \phantom{x} \\
    \sqrt{n+1} u_{n+1}(x) v_m(x) - \phantom{x} \\
    \sqrt{m+1} u_{n}(x) v_{m+1}(x),
\end{multline}
for $m \geq n$, while for $m < n$ we use the symmetry
$f_{n,m}(x) = f_{m,n}(x)$. On one side, for the ``regular''
solutions $u_n(x)$ one can use the following recurrence
relation:
\begin{equation} \label{eq:u-rec}
\begin{cases}
  & u_0(x) = e^{-x^2}, \\
  & u_1(x) = 2x e^{-x^2}, \\
  & u_n(x) = \frac{1}{\sqrt{n}}
    \pp*{2x u_{n-1}(x) -
      \sqrt{n-1} u_{n-2}(x)}.
\end{cases}
\end{equation}
On the other side, for the irregular wavefunctions $\phi_m$,
we have to setup a more careful construction, as explained
in~\cite{leorichter}. In the region given by the
Bohr--Sommerfeld radius, the classically allowed region, we
should employ a backward recursion instead. A safe choice
for this is to consider the region
\begin{align} \label{eq:safe-region}
  \abs{x} < \alpha_{4k} -
    \frac{1}{2}\frac{1}{\sqrt[3]{\alpha_{4k}}},
\end{align}
with $\alpha_n = \sqrt{n+1/2}$. In this case, it is
recommended to use the backward recursion as follows:
\begin{align} \label{eq:v-rec-b}
  v_m(x) = \frac{1}{\sqrt{m+1}}
    \pp*{2x v_{m+1}(x) - \sqrt{m+2} v_{m+2}(x)},
\end{align}
with initial values obtained from a semiclassical
approximation~\cite{leorichter} valid for large $m$
\begin{align}
  & v_m(x) = \kappa_m \deq
    \frac{(8\pi)^{1/4}}{\sqrt{\alpha_m \sin \tau_m}}
    \sin\pp*{\frac{1}{2} \alpha_m^2 \chi_m +
      \frac{\pi}{4}}, \\
\intertext{where}
  & \tau_m = \arccos(x/\alpha_m), \quad
    \chi_m = \sin(2\tau_m) - 2\tau_m. \nonumber
\end{align}
When $x$ is outside of the region \cref{eq:safe-region}, one
can employ the asymptotic form for $v_m$, which is
implemented by a forward recursion~\cite{leorichter}:
\begin{equation} \label{eq:v-rec-f}
\begin{cases}
  & v_0(x) = \frac{1}{x} e^{x^2}, \\
  & v_m(x) = \frac{\sqrt{m}}{2x} v_{m-1}(x).
\end{cases}
\end{equation}

\section{Recurrences for the Wigner function}
\label{sec:recur-wigner}

We have seen an optimized form of the reconstruction method
for the density matrix, but it may still be interesting to
recover also the Wigner function. \togli{The formula in
\cref{eq:wigner-lambda} does not perform well for lower
precision data types as there are several quantities that
become very large quite fast. Although one can mitigate this
by, for instance, computing first the logarithm of the
factorials, this can still be not enough in some cases.} We
present here two recursive methods to compute the functions
$\lambda_{n,d}$, leveraging the recurrence relations of the
Laguerre polynomials $L_n^d(x)$.

\subsection{Method 1}

Let us first recall the recursive definition of the
(generalized) Laguerre
polynomials~\cite{abramowitz1988handbook}:
\begin{align}
  & \begin{aligned}
  & L_0^d(x) = 1, \\
  & L_1^d(x) = 1 + d - x, \\
  & L_n^d(x) = a_{n,d}L_{n-1}^d(x)-b_{n,d}L_{n-2}^d(x), \\
  \end{aligned} \label{eq:lag} \\
\intertext{with}
  & a_{n,d} = \frac{2n+d-x-1}{n}, \quad
    b_{n,d} = \frac{n+d-1}{n}. \nonumber
\end{align}
We introduce the function
\begin{align}
  & \cG_n^d(x) = \sqrt{\frac{n!}{(n+d)!}} \cL_n^d(x),
\end{align}
where $\cL_n^d(x) = z(x)L_n^d(x)$ and $z(x) = \Lambda_0
e^{-x^2}$, with $\Lambda_0 = 4/\pi$. The function
$\lambda_{n,d}(x)$ in \cref{eq:wigner-numerics}, thus, can
be computed as
\begin{align}
  \lambda_{n,d}(x) = x^{d/2} \cG_n^d(x).
\end{align}
In the following we will drop the dependence on the variable
$x$. $\lambda_{n,d}$ can be viewed as a matrix of which only
the elements $n \leq M - d - 1$ have to be computed:
\begin{equation*}
\begin{array}{ccccc}
  \square & \square & \square & \square & \square \\
  \square & \square & \square & \square & \blacksquare \\
  \square & \square & \square & \blacksquare & \cdot \\
  \square & \square & \blacksquare & \cdot & \cdot \\
  \square & \blacksquare & \cdot & \cdot & \cdot
\end{array}
\end{equation*}
The starting point is $\lambda_{0,0} = z$, and the
first 2 rows can be obtained by
\begin{align}
  \lambda_{0,d} = \sqrt\frac{x}{d} \lambda_{0,d-1}, \quad
  \lambda_{1,d} = \frac{1+d-x}{\sqrt{d+1}}
    \lambda_{0,d},
\end{align}
while the first column $\lambda_{n,0} = \cL_n^0$ is just
given by the (scaled) Laguerre polynomials.
\begin{equation*}
  \begin{array}{ccccc}
    z & \sqrt{x}z & \lambda_{0,2} & \cdots & \cdots \\
    (1 - x) z & \frac{\sqrt{x}z}{\sqrt{2}}(2 - x)& \lambda_{1,2}& \cdots & \blacksquare \\
    \cL_2^0 & \square & \square & \blacksquare & \cdot \\
    \vdots & \square & \blacksquare & \cdot & \cdot \\
    \vdots & \blacksquare & \cdot & \cdot & \cdot
  \end{array}
\end{equation*}

At this point we can provide the general recursive formula,
valid for $n\geq 2$ and $d \geq 0$ (for $d=0$ the following
relation just simplifies to that of the Laguerre
polynomials, so that the first column can be computed with
specialized formulae):
\begin{align}
  & \lambda_{n,d} =
    a'_{n,d} \lambda_{n-1,d} -
      b'_{n,d} \lambda_{n-2,d}, \label{eq:lambda} \\
\shortintertext{where}
  & a'_{n,d} = \frac{2n+d-x-1}{\sqrt{n(n+d)}}, \nonumber \\
  & b'_{n,d} = \sqrt{\frac{(n-1)(n+d-1)}{n(n+d)}}. \nonumber
\end{align}

\medskip

\subsection{Method 2}

We can provide another implementation of the
\cref{eq:wigner-numerics} computing the function
$\lambda_{n,d}$ with different recurrence relations which
rely on the summation property of the Laguerre polynomials
$L_n^d$ which extends to the functions $\cG_n^d(x)$:
\begin{align}
  L_n^d = L_{n - 1}^d + L_n^{d - 1}.
\end{align}
Again the starting point of the recursion is $\lambda_{0,0}
= z$ and the first row is given by $\lambda_{0,d} =
\sqrt\frac{x}{d} \lambda_{0,d-1}$, while the first column is
obtained as before from $\cL_n^0$:
\begin{equation*}
  \begin{array}{ccccc}
    z & \lambda_{0,1} & \cdots & \cdots & \cdots \\
    (1 - x) z & \lambda_{1,1} & \cdots & \cdots & \blacksquare \\
    \cL_2^0 & \square & \square & \blacksquare & \cdot \\
    \vdots & \square & \blacksquare & \cdot & \cdot \\
    \vdots & \blacksquare & \cdot & \cdot & \cdot
  \end{array}
\end{equation*}

The general recursion formula, valid for $n\geq 1$ and
$d\geq 1$, is now given by:
\begin{align} \label{eq:lambda-2}
  & \lambda_{n,d} =
    \frac{\sqrt{n}\lambda_{n-1,d} -
      \sqrt{x}\lambda_{n,d-1}}{\sqrt{n+d}}.
\end{align}
With respect to the first method, in this case one needs an
additional vector to store the values $\lambda_{n,d-1}$.

\vspace{1cm}

\bibliography{bibliography}

\end{document}